\documentclass[prb,twocolumn,showpacs, amsmath, amssymb]{revtex4}
\allowdisplaybreaks
\usepackage{epsfig}
\usepackage{graphicx}

\newcommand{\beq} {\begin{equation}}
\newcommand{\eeq} {\end{equation}}
\newcommand{\beqa} {\begin{eqnarray}}
\newcommand{\eeqa} {\end{eqnarray}}

\begin{document}
\title{Host-Parasite Co-evolution and Optimal Mutation Rates for Semi-conservative Quasispecies}
\author{Yisroel Brumer and Eugene I. Shakhnovich}
\affiliation{Harvard University, 12 Oxford Street, Cambridge, Massachusetts 02138}
\date{\today}
\begin{abstract}
In this paper, we extend a model of host-parasite co-evolution
to incorporate the semi-conservative nature of DNA replication for both the
host and the parasite. We find that the optimal mutation rate for the 
semi-conservative and conservative hosts converge for realistic genome lengths,
thus maintaining the admirable agreement between theory and experiment found
previously for the conservative model and justifying the conservative approximation
in some cases. We demonstrate that, while the optimal mutation rate for a
conservative and semi-conservative parasite interacting with a given immune
system is similar to that of a conservative parasite, the properties away from this
optimum differ significantly. We suspect that this difference, coupled with 
the requirement
that a parasite optimize survival in a range of viable hosts, may help explain
why semi-conservative viruses are known to have significantly lower
mutation rates than their conservative counterparts.

\end{abstract}
\pacs{87.14Gg, 87.23Kg, 87.10+e}
\maketitle

\section{Introduction}

Introduced over 30 years ago, the quasispecies model of evolution \cite{Eigen1, Eigen} has provided
an invaluable tool for the study of complex evolutionary behaviors. In the model, a fitness 
landscape is introduced, which accounts, often in a highly approximate manner, 
for the complex interplay between genotype, phenotype and environment by assigning
a relative fitness for each genomic sequence (and thus associating phenotype 
with genotype, an approximation that must be treated with care). Through
the consideration of numerous individually mutating copies of a genome,
evolutionary systems can be studied analytically and numerically on
these fitness landscapes, which has provided enormous insight into
the process of evolution and the nature of mutation rates in real 
biological systems. In particular, it was found that a phase transition (known as the ``error catastrophe'')
occurs as the mutation rate increases, and a marked crossover can be observed
from the existence
of a quasispecies (wherein most individuals in the population contain genomes
close to a fitness peak) to a near-random walk in genome space with no
discernible quasispecies present \cite{Eigen}. 

The vast majority of the literature 
on the quasispecies model involve studies of asymptotic behavior on numerous 
stationary landscapes \cite{Galluccio, Schuster, Tarazona, Peliti}. 
This corresponds to a situation where static environmental conditions are
considered to be the dominant evolutionary pressure on a species. However, 
this picture fails to describe the cornucopia of evolutionary pressures in 
nature. Many organisms, parasites, survive through the detrimental use
of host biochemical processes. The parasite requires the host to live. The host
survives better if it can avoid or destroy the parasite, providing an
intriguing scenario: the host must evolve to defeat the parasite and the
parasite must evolve to evade the host's defenses. This creates a non-linear
feedback cycle as both species scour a time-dependent fitness landscape
that changes as the other species mutates.

Parasites are ubiquitous in nature, ranging from the microscopic (e.g. viruses,
bacteria, protozoa) to fungi, helminths and arthropods. The interaction between
parasites and hosts are very complex, with parasites exhibiting multistage
life cycles, inert phases, and the use of multiple intermediate hosts,
while hosts employ a wide variety of behavioral and immune defenses. This
ongoing struggle has been well documented in mammals, birds,
fish, bacteria and other organisms.

Recent work on time-dependent quasispecies landscapes \cite{timedep, timedep2}
has allowed for the study of a simple model of co-evolution by Kamp and 
Bornholdt \cite{Kamp, Kamp2}, discussed in detail in Section III. 
They derived a parameter-independent expression for the optimal mutation
rate for a host genome, which compared admirably with experimental results
on B-cell mutation rates \cite{Kamp}. An expression was also derived for 
optimal viral mutation rates \cite{Kamp2} which, although dependent on the parameters of
the model, explained numerous phenomena including the constancy of 
mutation rates within a viral class. However, this model considers the
interaction only between a conservatively replicating parasite and host.

In its conservative formulation, the quasispecies model considers  
single stranded genomes that produce multiple copies of itself,
each possessing a set of point mutations, while the original genome is
conserved.
 While this model is obviously applicable to numerous RNA-based 
viruses, the vast majority of organisms, including many viruses and other 
parasites, store genetic information in double stranded DNA. DNA replicates 
semi-conservatively through a series of steps discussed in Section II.
In a recent work, Tannenbaum et. al. \cite{Manny} reformulated
the quasispecies model to accurately represent semi-conservative systems,
which were found to display fundamentally different behavior than conservative
systems with respect
to the error catastrophe
in the infinite
time limit on a static landscape. Thus, to properly model the 
co-evolution of a parasite and its host, the host system must replicate 
semi-conservatively, while the parasite can be modeled as either conservative,
as in the case of many riboviruses, or semi-conservative, as by many lysogenic
double stranded DNA viruses or higher parasites. Retroviruses, such as HIV, 
likely display characteristics of both modes of replication, as do immune 
systems that undergo somatic hypermutation.

In this paper, we extend Kamp and
Bornholdt's model of co-evolution to the case of a semi-conservative host 
interacting with either a conservative or semi-conservative parasite. We consider
the optimal behavior for both the host and parasite, and demonstrate the 
similarities and differences between the conservative and semi-conservative
models.

The paper is organized as follows: in Section II we present the quasispecies
model and its extension to semi-conservative replication. In Section III we 
discuss the model of host-parasite co-evolution for both conservative and
semi-conservative organisms.
Section IV presents the results and discussion and Section V presents our conclusions. 

\vskip .2truein
\section{The Quasispecies Model}

In this section, we present some necessary background on the conservative and 
semi-conservative quasispecies models for the purpose of a self-contained 
discussion. Greater detail may be found in the original papers.  

\subsection{Conservative Replication}

The quasispecies model studies the evolution of a population of organisms, 
each with a genome $ \phi = s_1s_2\cdots s_n $, 
where each $ s_i $ represents a ``letter'' chosen from an alphabet of size $ S $. 
Often, $ S $ is chosen to be two to model the 
pyrimidine and purine groups or four to model the nucleotides. Assuming 
first-order growth kinetics and associating phenotype with genotype (i.e. that the
growth rate of an individual is directly determined by $ \phi $), it
can be shown that 
\begin{equation}
\frac{dx_{\phi}}{dt} = \sum_{\phi'}A(\phi')W(\phi, \phi')x_{\phi'} - f(t)x_{\phi},
\end{equation}
where $ x_{\phi} $ denotes the fraction of the population with genome $\phi$, 
$ A(\phi) $ represents the fitness, or growth rate, of sequence $\phi$,
$W(\phi, \phi')$ is the likelihood of creating sequence $\phi$ from
$\phi'$ by mutations, and $ f(t) = \sum_{\phi}A(\phi)x_{\phi} $ is the average fitness of the 
population, holding the population size constant and introducing competition.
If only point mutations are allowed and a genome-independent 
mutation probability $\epsilon$ is assumed, then $ W(\phi, \phi') $
can be written in terms of the genome length $n$ and the number of bases at which $\phi$ and $\phi'$ 
differ, the Hamming distance $ HD(\phi, \phi')$, as
\begin{equation}
W(\phi, \phi') = (\frac{\epsilon}{S - 1})^{HD(\phi, \phi')}(1 - \epsilon)^{n - HD(\phi, \phi')}.
\end{equation}
These equations can be greatly simplified in the case of a single fitness peak
landscape, where a master sequence, $\phi_0$, has a fitness much 
greater than all other sequences. The rest of the genomes are assumed to be 
equally fit, which can be described by the growth rates
\begin{equation}
A(\phi) = \left\{ \begin{array}{cc} 
             \eta & \phi \neq \phi_0 \\
             \sigma \gg \eta & \phi = \phi_0
             \end{array} 
             \right. . 
\end{equation}
The sequences can then be grouped into Hamming classes based on their distance from
the master sequence by defining
\begin{equation}
 w_l = \sum_{\phi \in \{\phi \mid HD(\phi, \phi_0) = l\}} x_\phi 
\end{equation}
and
\begin{equation}
 \begin{array}{cc}
 A(l) \equiv A(\phi) & \phi \in  \{\phi \mid HD(\phi, \phi_0) = l\}.
 \end{array}
\end{equation}
This reduces the problem from $ S^n $ dimensions to $ n + 1 $ dimensions. 
If mutations that lead from higher to lower Hamming distances are ignored
(an approximation that becomes exact as $ n \rightarrow \infty$), 
\begin{eqnarray}\nonumber
&&\frac{dw_l}{dt} = \sum_{l' = 0}^{l} \frac{(n - l')!}{(n - l)!}A(l')(\frac{\epsilon}{S-1})^{l-l'}(1-\epsilon)^{n - (l - l')}w_{l'} \\[0.25in] && - f(t)w_l,
\end{eqnarray}
where $ f(t) = \sum_{l}A(l)w_{l} = \sigma w_0 + \eta(1 - w_0) = (\sigma -
\eta)w_0 + \eta $. Defining $ y_i = w_i \exp(\int_0^t f(s)ds) $ 
removes the non-linearity in these equations and the linear set of differential
equations can be solved for any Hamming class.
The solution for the master sequence is 
\begin{equation}
y_0(t) = y_0(0)e^{q^n\sigma t}
\end{equation}
and, for the first Hamming class,
\begin{equation}
y_1(t) = y_0(0) n\left( \frac{(e^{q^n\sigma t} - e^{q^n \eta t})(1 - q)\sigma}{(S-1)(\sigma - \eta)q}\right),
\end{equation}
where $ q = 1 - \epsilon $, a definition we shall use throughout the paper.

\subsection{Semi-conservative Replication}

In order to properly model a semi-conservative system,  
a double stranded molecule generated from an alphabet of size $ S $ must
be considered,
where each ``letter'' $ i $ uniquely pairs with $ (i + S/2)\mod S $. DNA 
requires $ S = 4 $, where the letters can be assigned as $ A \equiv 1, 
G \equiv 2, T \equiv 3, C \equiv 4 $. A single DNA molecule of length n consists
of a strand $ \phi = s_1 s_2 \cdots s_n $ and a complementary strand $\underline{\phi}
 = \underline{s}_1\underline{s}_2 \cdots \underline{s}_n $ where $ \underline{s}_i $ denotes the
complement of $ s_i $. Hence, each DNA molecule may be represented by the pair
$ \{\phi, \underline{\phi}\} \equiv \{\underline{\phi}, \phi\} $.

\begin{figure}[ht]
\includegraphics[width=1\linewidth]{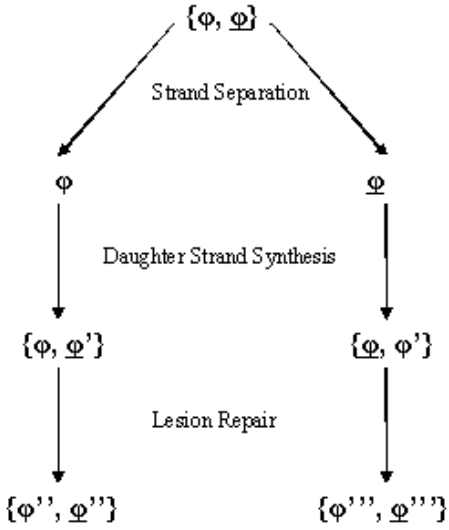}
\caption{A schematic model of DNA replication. Adapted from Tannenbaum {\it et. al.} \cite{Manny}.}
\end{figure}
When a semi-conservative molecule replicates, it undergoes a three step 
process shown schematically in Fig. 1. First, each genome $\{\phi, \underline{\phi}\}$
unzips to form two single stranded genomes, $\phi$ and $\underline{\phi}$. Each strand
is then copied to produce two new pairs, $ \{\phi, \underline{\phi}'\} $ and $ \{\underline{\phi}, 
\phi'\}$, where the primes denote the fact that the two fresh strands may contain 
replication errors. At this point, proofreading mechanisms can
distinguish between the new and old strands and may fix all or some of the 
replication errors, which can be spotted by the fact that 
$ s_i' \neq \underline{s}_i$. All of these repair mechanisms are included in the 
base-independent error probability $ \epsilon $. In the last step, the new
and old strands become indistinguishable. Various maintenance enzymes repair 
the remaining mismatches, but cannot determine which of the strands $\phi$ and
$ \underline{\phi}' $ is the newly replicated strand. Hence, the repair is made in the
new strand with 50\% probability and in the old strand with 50\% probability.
The final result is that the original strand $ \{\phi, \underline{\phi}\} $ is replicated
to create two new strands, $ \{\phi'', \underline{\phi}''\} $ and $ \{\phi''', \underline{\phi}'''\} $ \cite{Voet}. 

The quasispecies equations for this system can be
written as \cite{Manny}
\begin{eqnarray} \nonumber
\frac{dx_{\{\phi, \underline{\phi}\}}}{dt} = \sum_{\{\phi', \underline{\phi}'\}}A(\{\phi', \underline{\phi}'\})x_{\{\phi', \underline{\phi}'\}}(p(\phi',  \{\phi, \underline{\phi}\}) +\\
p(\underline{\phi}',  \{\phi, \underline{\phi}\})) - (A(\{\phi, \underline{\phi}\}) + f(t))x_{\{\phi, \underline{\phi}\}},
\end{eqnarray}
where $ f(t) = \sum_{\phi}A(\{\phi, \underline{\phi}\})x_{\{\phi, \underline{\phi}\}} $
and $ p(\phi',  \{\phi, \underline{\phi}\}) $ represents the probability that
the unzipped strand $ \phi' $ will produce the pair $ \{\phi, \underline{\phi}\} $.
To make these equations more useful, we can define $ A(\phi)  \equiv A(\phi, 
\underline{\phi}) $ and $x_{\phi} \equiv \frac{1}{2}x_{\{\phi, \underline{\phi}\}} $
if $ \phi \neq \underline{\phi} $ and $x_{\phi} \equiv x_{\{\phi, \underline{\phi}\}} $ if
$ \phi = \underline{\phi} $. After some manipulation, we obtain 
\begin{eqnarray} \nonumber
&&\frac{dx_\phi}{dt} = \\ [0.25in] \nonumber
&&2\sum_{\phi'}A(\phi')x_{\phi'}(\frac{\epsilon/2}{S-1})^{HD(\phi, \phi')}(1-\frac{\epsilon}{2})^{n - HD(\phi, \phi')} \\ [0.25in]
&& - (A(\phi) + f(t))x_\phi,
\end{eqnarray}
where $ f(t) = \sum_\phi A(\phi)x_\phi.$

We now turn our attention to semi-conservative replication on a single 
fitness peak landscape. This case is more complicated than for a conservative
system, since viability genes often exist on both strands in nature. Hence, if there exists a sequence $\phi_0$ with fitness $\sigma$, it 
stands to reason that the sequence $\underline{\phi}_0$ should have fitness $\sigma$
as well, effectively creating a double fitness peak landscape (this assumption
is by no means fundamental to the work). However, 
noting that $x_{\phi} = x_{\underline{\phi}}$ for all times, both by definition and
by conservation in Equation (10), this difficulty can be sidestepped. As long
as $ n$ is not too small, the area around each fitness peak can be locally 
treated as a single fitness peak landscape as the two peaks are distant in 
sequence space. Hence, ignoring back mutations,  the two master
sequences obey the equations
\begin{eqnarray} 
\nonumber
&& \frac{dw_0}{dt} = 2(1-\epsilon/2)^n \sigma \underline{w}_0 - (\sigma + f(t))w_0 =\\ [0.25in] 
&& 2(1-\epsilon/2)^n \sigma w_0 - (\sigma + f(t))w_0 \\[0.25in] 
\nonumber
&& \frac{d\underline{w}_0}{dt} = 2(1-\epsilon/2)^n \sigma w_0 - (\sigma + f(t))\underline{w}_0 = \\ [0.25in]
&& 2(1-\epsilon/2)^n \sigma \underline{w}_0 - (\sigma + f(t))\underline{w}_0,
\end{eqnarray}
where $ w_i $ represents the concentration of the $i$th Hamming class as before.
Therefore, we can re-define the concentration of the master sequence to include both 
$ w_0 $ and $ \underline{w}_0 $ and use equation (11) for the sum of the two. While this is not
strictly necessary and has no effect on the results, it does reduce the 
bookkeeping, and the characteristics of the individual peaks can be obtained
by simply dividing by two.
A similar procedure yields 
\begin{eqnarray} \nonumber
&& \frac{dw_1}{dt} = 2(1-\epsilon/2)^{n-1}(\frac{\epsilon/2}{S-1}) n \sigma w_0 + \\[0.25in]
&& 2(1-\epsilon/2)^{n} \eta w_1 - (\eta + f(t))w_1,
\end{eqnarray}
where we include sequences of Hamming distance one away from both master sequences.
The definition  $ y_i = w_i \exp(\int_0^t f(s)ds) $ once again removes the 
non-linearity. The solutions for the first two Hamming classes are
\begin{eqnarray}
&& y_0(t) = y_0(0)e^{2\sigma(1-\epsilon/2)^{n} - \sigma} \\ [0.25in] \nonumber
&& y_1(t) = y_0(0) n \left( \frac{ \sigma \epsilon (1- \epsilon/2)^{n - 1}} 
 {(S-1)(\sigma - \eta)(2(1- \epsilon/2)^n - 1)} \right) \times\\ [0.25in]
&& (e^{\sigma(2(1- \epsilon/2)^n - 1)t} - e^{\eta(2(1- \epsilon/2)^n - 1)t})
\end{eqnarray}

\vskip .2truein
\section{Host-Parasite Co-evolution}

Historically, the main focus of research on the quasispecies model has related 
to static and equilibrium properties of the system \cite{Tarazona, Galluccio,
Pastor,  Franz, Campos, Alves, Altmeyer}. A number of recent works, however, 
have explored
the dynamics of the system under various conditions \cite{timedep, timedep2, 
timedep3, Wilke}, which has allowed the study of the simple model
of co-evolution described here. Following the work of Kamp and Bornholdt\cite{Kamp, Kamp2},  
we envision a 
population of host and parasite organisms (which we shall refer to as the
immune system and virus), each described by a set of
quasispecies equations. Ignoring the interspecies interaction, the immune 
and viral genomes, of length $ n_{is} $ 
and $ n_v $, respectively, evolve independently on a single fitness peak landscape, where
the master sequences have fitness $ \sigma_{is} \gg \eta_{is} $ and 
$\sigma_{v} \gg \eta_{v} $. To model the deleterious effect of the immune
system on the virus, the dominant immune genome imposes a large death rate 
$ \delta $ on the corresponding viral sequence. If this dominant immune genome
matches the viral master sequence, the viral fitness peak will move to an
arbitrary sequence of the first Hamming class. The viral quasispecies then
adapts to this new fitness peak on a timescale $\tau_v$, the time required for
the population of the new master sequence to overtake that of the old. At this
point, the immune system fitness peak adjusts to match the new viral peak,
and adapts on a similarly defined timescale $\tau_{is}$. Thus, through the
iteration of these steps, the viral
fitness peak scours sequence space in an attempt to avoid the immune system, which
follows on its heels. Applying recent results on dynamic fitness landscapes \cite{timedep}, regions of 
stability can be defined for both the viral and immune quasispecies
by determining a characteristic timescale for regrowth of a new master sequence. 
If the landscape moves slowly enough, the master sequence has time to 
regenerate to the master sequence concentrations reached before the peak shift
and the species will survive for all time. If, however, the master sequence
cannot regenerate rapidly enough, a second peak shift will occur before the
new master sequence reaches the concentration held by the old master 
sequence before the first shift. The third master sequence cannot reach the 
levels
of the second, and this continues until, eventually, there is no discernible
master sequence in the population.
For the conservative case, this can be stated rigorously by comparing the 
growth of a single member of the first Hamming class described by equation 
(8) with $ e^{\eta \tau} $, 
the uninhibited growth of a random sequence far from the fitness peak 
(as mutations in and out of this sequence should cancel). Using equation (8)
this ratio can be
defined, for both the immune and viral quasispecies, as
\cite{Kamp, timedep2}
\begin{eqnarray} \nonumber
&& \kappa \equiv \frac{w_1(\tau)}{n e^{\eta \tau} w_0(0)} \\[0.25in]
&& \equiv \left( \frac{(e^{(q^n\sigma -\eta)\tau} - e^{(q^n\eta -\eta)  \tau})(1 - q)\sigma}{(S-1)(\sigma - \eta)q}\right),
\end{eqnarray}
where $\tau$ is the lag time between peak shifts and the parameters $ \{q, \sigma, \cdots\} $ represent the parameters for either species. The quasispecies
survives only when  $\kappa \geq 1$.

The last piece necessary to complete the co-evolution model, then, is the speed
with which the landscape moves. By the definition of our model, $\tau$
is the sum of the the time required for the regeneration of the virus, 
$\tau_v $, plus the time required for the regeneration of the immune system,
$ \tau_{is} $. Hence, we must solve for $ \tau = \tau_{is} + \tau_v$ where
\begin{eqnarray}
e^{(q_v^n \eta_v - \delta)\tau_v}w_{0,v}(\tau) = e^{(q_v^n\sigma_v \tau_v)}w_{1,v}(\tau)/n\\[0.25in]
e^{q_{is}^n \eta_{is}\tau_{is}}w_{0, is}(\tau) = e^{(q_{is}^n\sigma_{is} \tau_{is})}w_{1, is}(\tau)/n.
\end{eqnarray}
This can be solved to obtain
\begin{eqnarray} \nonumber
&& e^{(q_v^n \eta_v - \delta)\tau_v}e^{q_v^n\sigma_v \tau} = \\[0.25in]
&&e^{q_v^n\sigma_v \tau_v} \frac{(e^{q_v^n\sigma_v\tau} - e^{q_v^n\eta_v  \tau})(1 - q_v)\sigma_v}{(S-1)(\sigma_v - \eta_v)q_v} \\ [0.25in] \nonumber
&&e^{q_{is}^n \eta_{is}\tau_{is}}e^{q_{is}^n\sigma_{is} \tau} = \\[0.25in]
&&e^{q_{is}^n\sigma_{is} \tau_{is}} \frac{(e^{q_{is}^n\sigma_{is}\tau} - e^{q_{is}^n\eta_{is}  \tau})(1 - q_{is})\sigma_{is}}{(S-1)(\sigma_{is} - \eta_{is})q_{is}},
\end{eqnarray}
which yields, with the reasonable approximations that $q \approx 1$ and
$ \sigma \gg \eta$ (the latter of which is used throughout the paper),
\begin{eqnarray}
&& \tau_v \cong -\frac{\ln(\frac{1-q_v}{S-1})}{q_v^n(\sigma_v - \eta_v) + \delta} \\[0.25in]
&& \tau_{is} \cong -\frac{\ln(\frac{1-q_{is}}{S-1})}{q_{is}^n(\sigma_{is} - \eta_{is})}
\end{eqnarray}
These equations can be applied to determine the optimal mutation rate for
both the host and the parasite. The host can minimize the region of viability
for the parasite by evolving a mutation rate such that
\begin{eqnarray}
\frac{\partial \kappa_v}{\partial \epsilon_{is}} = 0,
\end{eqnarray}
yielding \cite{Kamp}
\begin{eqnarray}
\epsilon_{is} - 1 - n_{is}\epsilon_{is}\ln\left(\frac{\epsilon_{is}}{S-1}\right)= 0.
\end{eqnarray} 
This equation has the nice quality of being independent of the parameters of the
immune model, as well as the properties of the virus. The solution to this equation
is shown in Fig. 2 and compared to the experimentally verified mutation rate
for human B-cell receptors. 
This is discussed at length in section IV.

Optimizing the viral mutation rate requires solving for
\begin{eqnarray}
\frac{\partial \kappa_v}{\partial \epsilon_v} = 0
\end{eqnarray}
or \cite{Kamp2}
\begin{widetext}
\begin{eqnarray} \nonumber
&& (q_v^{n_v}(\sigma_v - \eta_v) + \delta)(n_v(q_v - 1)q_v^{2n_v} \sigma_v^2 \tau_{is} 
 + \delta(q_v + (q_v - 1)n_vq_v^{n_v} \sigma_v \tau_{is}) 
 + \tau_v(q_v - q_v^{n_v + 1} - (q_v - 1)n_vq_v^{2n_v}\sigma_v\tau_{is})) \\[0.25in]
&& + n_vq_v^{n_v}(q_v - 1)(\eta_v^2 - \delta \sigma_v - \eta_v\sigma_v)ln(\frac{1-q_v}{S-1}) = 0,
\end{eqnarray}
\end{widetext}
the solution of which is shown in Fig. 3 for a chosen set of parameters.

We now turn our attention to the central theme of this paper, the co-evolution
of semi-conservative organisms. Applying the results of section II and following
the procedure outlined above, we find, for a semi-conservatively replicating 
host,
\begin{eqnarray} \nonumber
&& \kappa_{is} = \left(\frac{\sigma_{is} \epsilon_{is} (1 - \epsilon_{is}/2)^{n_{is} - 1}}{(S-1)(\sigma_{is} - \eta_{is})(2(1- \epsilon_{is}/2)^{n_{is}} - 1)}\right) \times \\[0.25in] \nonumber
&& (e^{(2\sigma_{is}(1-\epsilon_{is}/2)^{n_{is}} - \sigma_{is} - \eta_{is})\tau} - \\ [0.25in]
&& e^{(2\eta_{is}(1 - \epsilon_{is}/2)^{n_{is}} - 2\eta_{is})\tau}) \\ [0.25in]
&& \tau = \tau_{is} + \tau_{v} \\[0.25in]
&& \tau_{is} = -\frac{\ln \left( \frac{(1-\epsilon_{is}/2)^{n_{is}}\epsilon_{is}}
{(2(1-\epsilon_{is}/2)^{n_{is}} - 1)(S-1)}\right)}
{(2(1-\epsilon_{is}/2)^{n_{is}} - 1)(\sigma_{is} - \eta_{is})}. 
\end{eqnarray}
A conservatively replicating virus interacting with this host will still
follow the behavior described by equations 16 and 21, albeit with
the proper, semi-conservative $\tau_{is} $ defined above. For the case of
a semi-conservative virus we obtain
\begin{eqnarray} \nonumber
&& \kappa_{v} = \left(\frac{\sigma_{v} \epsilon_{v} (1 - \epsilon_{v}/2)^{n_{v} - 1}}{(S-1)(\sigma_{v} - \eta_{v})(2(1- \epsilon_{v}/2)^{n_{v}} - 1)}\right) \times \\[0.25in] \nonumber
&& (e^{(2\sigma_{v}(1-\epsilon_{v}/2)^{n_v} - \sigma_{v} - \eta_v)\tau} - \\ \nonumber
\\ [0.25in] && e^{(2\eta_{v}(1 - \epsilon_{v}/2)^{n_{v}} - 2\eta_{v})\tau}) \\[0.25in] 
&& \tau_{v} = -\frac{\ln \left( \frac{(1-\epsilon_{v}/2)^{n_v}\epsilon_{v}}
{(2(1-\epsilon_v/2)^{n_{v}} - 1)(S-1)}\right)}
{(2(1-\epsilon_v/2)^{n_{v}} - 1)(\sigma_{v} - \eta_{v}) + \delta}. \label{}
\end{eqnarray}

\begin{figure}[ht]
\includegraphics[width=1\linewidth]{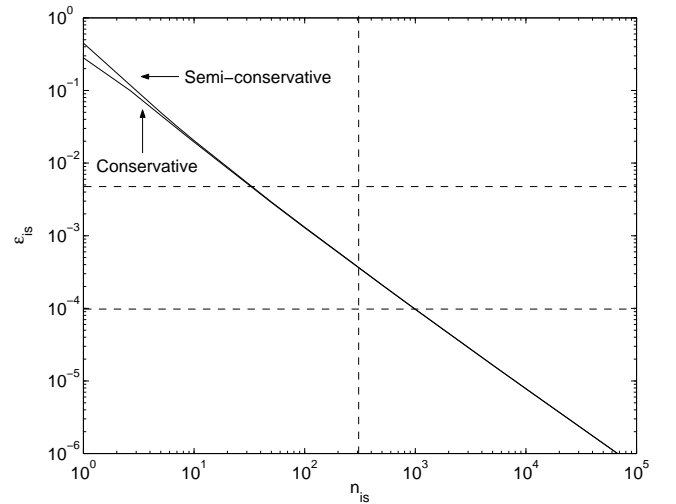}
\caption{Optimal immune system mutation rate vs. $n_{is}$. The dashed lines
represent experimental values for somatic hypermutation of B-cell 
complementary determining regions, adapted from \cite{Kamp}.}
\end{figure}
We now proceed to find the optimal mutation rates for both organisms.
Differentiating $\kappa_{v}$ by $\epsilon_{is} $ and setting the result to
zero gives us a criterion for the optimal immune mutation rate, 
\begin{widetext}
\begin{eqnarray}
\frac{-2 + \epsilon_{is} + n_{is}\epsilon_{is} - 2\,
     {\left( 1 - \frac{\epsilon_{is}}{2} \right) }^{n_{is}}\,
     \left( -2 + \epsilon_{is} + n_{is}\,\epsilon_{is}\,
        \ln (\frac{\left(S-1 
              \right) \,   
            \left( 2 - 
              {\left( 1 - \frac{\epsilon_{is}}{2} \right) }^{-n_{is}}
              \right) }{\epsilon_{is}}) \right) }{{\left( 1 - 
        2\,{\left( 1 - \frac{\epsilon_{is}}{2} \right) }^{n_{is}} \right) }^
     2\,\left(\epsilon_{is} - 2\right) \,\epsilon_{is}\,
    \left( \sigma_{is} - \eta_{is} \right) } = 0.
\end{eqnarray}
\end{widetext}
This equation has all of the nice properties of Equation (24), defining an 
optimal mutation rate for any genome length, independent of the parameters
of the system. The solution to this equation is plotted in Fig. 2, along with
the conservative solution and the experimental range for 
observed rates per base pair per generation of somatic hypermutation
in the complementary 
determining regions (CDR's) found in B-cell antigen receptors. 

\begin{figure}[ht]
\includegraphics[width=1\linewidth]{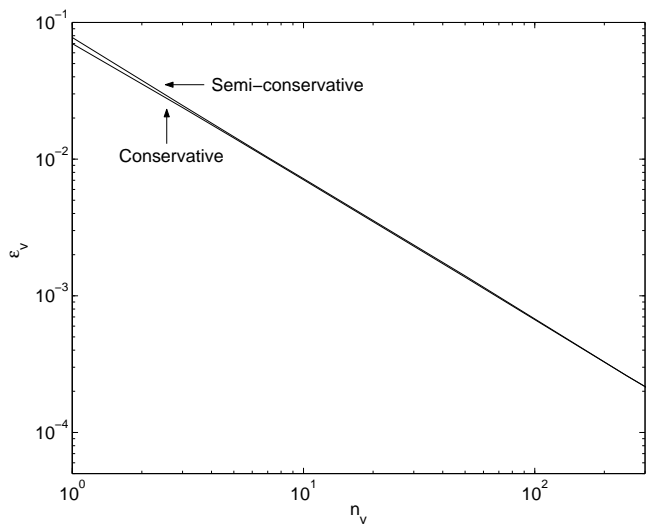}
\caption{Optimal viral mutation rate vs. $n_v$ for a conservative and semi-conservative virus
interacting with a semi-conservative immune system. $n_{is} =  100, \sigma_{is}= \sigma_v = 100, \eta_{is} = \eta_v = 1, \delta = 200, \epsilon_{is} = 0.001.$}
\end{figure}
To maximize the stability of the viral quasispecies we set $\partial \kappa_v / 
\partial \epsilon_v = 0 $ as before. After a fair bit of work, we obtain
an unwieldy expression omitted here in the interest of space \cite{expression}.
The expression simplifies immensely in the limit  $\delta \rightarrow 
\infty $, the limit of an ideally efficient immune system. In this limit, 
\begin{eqnarray} \nonumber
&&\frac{n_v\epsilon_v}
{2(1 - 2(1-\frac{\epsilon_v}{2})^{n_v})^2} + \\[0.25in] 
&&\frac{n_v\sigma_v\epsilon_v(1-\frac{\epsilon_v}{2})^{n_v}\tau_{is} - 1}{1 - 2(1-\frac{\epsilon_v}{2})^{n_v}}= 0.
\end{eqnarray}
The ideally efficient immune system is not an unreasonable approximation, as immune systems are highly efficient in 
destroying invaders once a suitable antibody is produced. 
The full expression as well as the above limiting form are dependent on both 
the parameters of the model 
and the properties of the immune system as in the conservative case. 
The solution of the full expression for a particular set of parameters is shown in Fig. 3.

\vskip .2truein
\section{Results and Discussion}

Given the fundamental differences between semi-conservative and conservative modes of 
replication, the most striking aspect of Fig. 2 and 3 is the similarity
between the conservative and semi-conservative optimal mutation rates at high
$ n$ , 
particularly for the viral species. This is most easily understood by
noting that, as $ (1 - \epsilon/2)^n \rightarrow 1$  for any semi-conservatively
\begin{figure}[ht]
\includegraphics[width=1\linewidth]{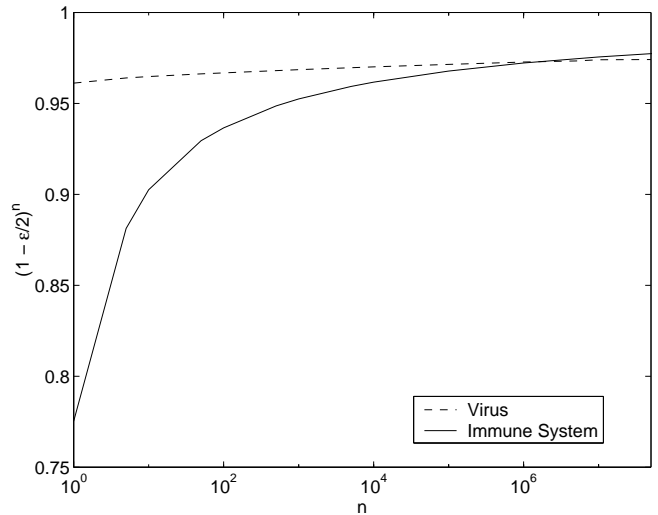}
\caption{$(1 - \epsilon/2)^n$ for the optimal mutation rate of a 
semi-conservative immune system and virus. This parameter can be used as
a measure of the ``conservativeness'' of a semi-conservative system.
$\sigma_{is}= \sigma_v = 100, \eta_{is} = \eta_v = 1, \delta = 200, \epsilon_{is} = 0.001.$}
\end{figure}
replicating organism, the probability that a mutation will be found in the 
{\em original} strands after replication vanishes. Hence, in this limit,
semi-conservative and conservative replication are expected to mimic
each other. This parameter is shown in Fig. 4 for the optimal viral and
immune mutation rates. Clearly, with the exception of small immune genomes,
the conservative system can be used as a good approximation for semi-conservative
replication. It is important to note, however, that this knowledge could 
not have been extracted from the data for the conservative system. A large
value for $ (1 - \epsilon/2)^n $ in the conservative system is a necessary
but not sufficient criterion to justify the use of a conservative model,
and the full semi-conservative calculation is required.

\begin{figure}[ht]
\includegraphics[width=1\linewidth]{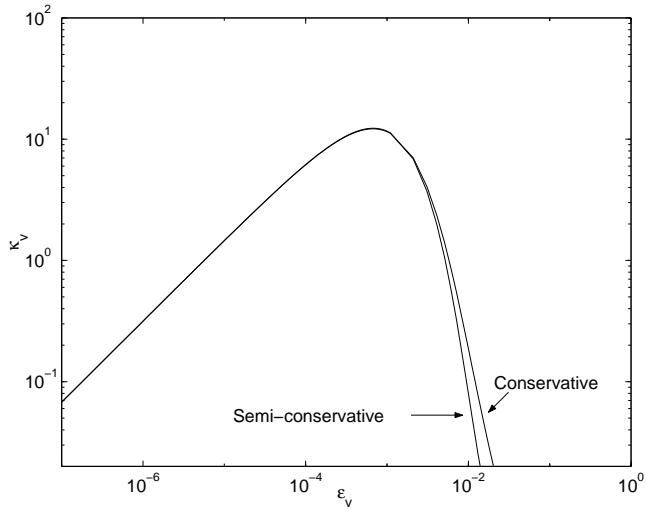}
\caption{$\kappa_v$ vs. $\epsilon_v$ for a conservative and semi-conservative virus
interacting with a semi-conservative immune system. $n_{is} = n_v = 100, \sigma_{is}= \sigma_v = 100, \eta_{is} = \eta_v = 1, \delta = 200, \epsilon_{is} = 0.001.$}
\end{figure}
Equation (24) remains dependent on the parameters of the model, but general
trends are obvious when biologically reasonable parameters are employed.
While the extremal behavior of Equation (27) and (30) differs little from
Equation (16) 
for genome lengths that are not too small, the behavior away from the maxima differs greatly. Fig.
5 displays $\kappa_v$ vs. $\epsilon_v$ for a given set of parameters for
both the conservative and semi-conservative models. It is immediately clear
that, while the two models coincide at small $ \epsilon $ (with a slightly
higher peak height for either species for some parameters), their behavior 
differs
greatly otherwise, with the semi-conservative model displaying a more 
drastic dropoff in viability as $ \epsilon $ increases, true for all 
biologically reasonable parameters studied. The parameters shown in Fig. 5 
were chosen as a representative, rather than extreme, example of this behavior.
The importance of this result
is best understood in light of the evolutionary pressures one would
expect a viral population to encounter. The independence 
of Equation (32) from the properties of the viral system suggests that there
exists an optimal mutation rate for an immune receptor {\em independent} of
the qualities of the parasite against which it is defending. Thus, it is 
reasonable to expect (within the limitations imposed by additional evolutionary
pressures, such as the need to distinguish between self and foreign antigens)
an immune receptor to evolve this mutation rate nearly exactly. However, in the viral
case, the optimal mutation rate depends strongly on the nature of the immune
system it is attacking. Thus, the virus must evolve the mutation rate that 
maximizes its overall viability against the range of immune systems it is likely
to infect, including both inter- and intra-species viability. 
The mutation 
rate that optimizes defense against one host may be a poor choice for another,
and the virus must find the mutation rate that affords the best protection
against all hosts, even if this is not the best mutation rate for evading
any particular immune system.
Such a compromise
clearly involves the behavior of $\kappa_v$ over a wide range of $ \epsilon$, 
rather than just at the maximum. One would therefore expect the more drastic
dropoff at higher $ \epsilon $ to force the semi-conservative virus
to develop a lower mutation rate so as to increase its viability against
immune systems that lower the $\epsilon_v$ with the maximal value of $\partial \kappa_v / \partial \epsilon_v $.
Quantifying this statement requires an intelligent estimate of the
distribution of immune properties, a subject of future research.
Qualitatively, this agrees well with the experimentally verified fact that 
semi-conservative viruses
display significantly lower mutation rates than their conservative counterparts
\cite{Drake, Drake2}. 

\section{Conclusions}

In this paper, we have extended Kamp and Bornholdt's model of co-evolution to 
incorporate the semi-conservative nature of DNA replication for both species. 
A parameter-independent expression was derived for the optimal mutation rate of
an immune receptor, which agrees well with experimental data. Convergence of the 
conservative and semi-conservative results was demonstrated for realistic genome
sizes, justifying the use of a conservative model in this case. 

Optimizing the stability of the immune species yielded a maximum 
that coincides with the conservative model for realistic genome sizes. 
A similar correspondence exists for the virus, albeit with a dependence on the parameters of the model.
Away from the maximum, the conservative and semi-conservative models
display different behaviors that provides a possible explanation
for the high mutation rates found in conservative viruses. It
is always dangerous to extrapolate from a simplified model of this kind to
the complex systems found in nature. A true virus and immune system must 
contend with innumerable evolutionary pressures, biological, chemical and
otherwise, such as the requirement that T-cells recognize and do not bind host
proteins. The work represented in this paper describes a generalized
model which we feel captures the robust qualitative features of host-parasite
coevolution, providing insight into the complex workings of nature.

\begin{acknowledgments}
The authors would like to thank Emmanuel Tannenbaum, Brian Dominy, Eric Deeds,
and Stefan Bornholdt for useful and interesting discussions.
\end{acknowledgments}

\bibliography{ConsImmune.bib}

\end{document}